%% file: main.tex
\documentclass{article}
\usepackage{spconf,amsmath,graphicx}
\usepackage{hyperref}
\usepackage{bbm}
\usepackage{todonotes}
\usepackage{multirow}
\usepackage{tabularx}
\usepackage{subfig}
\usepackage{booktabs}
\usepackage{enumitem}
\setlist{nosep, leftmargin=14pt}
\usepackage{mwe}

\DeclareMathOperator*{\median}{median}
\DeclareMathOperator{\proj}{proj}
\DeclareMathOperator{\HD}{HD}

\title{Scale-specific auxiliary multi-task contrastive learning for deep liver vessel segmentation}

\name{A. Sadikine $^{\bullet, \star}$
\qquad \hspace{-0.7cm} B. Badic $^{\bullet, \flat}$
\qquad \hspace{-0.7cm} J.-P. Tasu $^{\bullet, \circ}$
\qquad \hspace{-0.7cm} V. Noblet $^{\sharp}$
\qquad \hspace{-0.7cm} P. Ballet $^{\bullet, \star}$
\qquad \hspace{-0.7cm} D. Visvikis $^{\bullet}$
\qquad \hspace{-0.7cm} P.-H. Conze $^{\bullet, \diamond}$
\thanks{This work was partially funded by LaBeX CAMI (grant ANR-11-LABX-0004) and France Life Imaging (grant ANR-11-INBS-0006).}}

\address{$^{\bullet}$ LaTIM UMR 1101, Inserm, Brest, France 
$^{\star}$ University of Western Brittany, Brest, France \\
$^{\flat}$ University Hospital of Brest, Brest, France
$^{\circ}$ University Hospital of Poitiers, Poitiers, France \\
$^{\sharp}$ ICube UMR 7357, CNRS, Strasbourg, France
$^{\diamond}$ IMT Atlantique, Brest, France
}

\begin{document}
\maketitle

\begin{abstract}
Extracting hepatic vessels from abdominal images is of high interest for clinicians since it allows to divide the liver into functionally-independent Couinaud segments. In this respect, an automated liver blood vessel extraction is widely summoned. Despite the significant growth in performance of semantic segmentation methodologies, preserving the complex multi-scale geometry of main vessels and ramifications remains a major challenge. This paper provides a new deep supervised approach for vessel segmentation, with a strong focus on representations arising from the different scales inherent to the vascular tree geometry. In particular, we propose a new clustering technique to decompose the tree into various scale levels, from tiny to large vessels. Then, we extend standard 3D UNet to multi-task learning by incorporating scale-specific auxiliary tasks and contrastive learning to encourage the discrimination between scales in the shared representation. Promising results, depicted in several evaluation metrics, are revealed on the public 3D-IRCADb dataset.
\end{abstract}

\begin{keywords}
vascular segmentation, multi-scale geometry, auxiliary tasks, contrastive learning, tubular structures
\end{keywords}

\section{Introduction}
\label{sec:intro}

Probing liver vessels from abdominal images is paramount for diagnosis, therapy, and surgery planning intents. Since manually performing dense voxel-level annotations of vascular structures from Computed Tomography (CT) scans is complex, time-consuming, and prone to expert variability, there remains a renowned need for an automated computerized approach that could effectively capture tubular structures while managing low contrast, high signal-to-noise, intensity variations, complex bifurcation, and multi-scale geometry. Numerous studies have addressed the challenge of vessel segmentation through vesselness enhancement \cite{frangi1998multiscale}, active contours \cite{zeng2018automatic}, or tracking \cite{kaftan2009two} methods. With imaging data and computational resources availability, supervised semantic segmentation approaches founded on Convolutional Neural Networks (CNN) \cite{long2015fully, cciccek20163d} have significantly raised and promoted notable performance for miscellaneous tasks. Despite a good ability to extract visceral organ contours \cite{kavur2021chaos,conze2021abdominal}, their ability to delineate hepatic vascular systems still remains a challenge.

\input{fig1}

Typically, blood vessel extraction is tackled without taking advantage of deep Multi-Task Learning (MTL) \cite{vandenhende2021multi} whose aim is to improve generalization, learning efficiency and prediction accuracy by leveraging domain-specific information from related tasks. However, considering blood vessel segmentation with multiple (sub-)tasks appears promising. Further, the network architectures related to MTL are multifarious and adjusted depending on the specific concerns under investigation \cite{vandenhende2021multi}. In the setting of single-input and multi-output network, which is a component of encoder-focused architectures, the encoding stage shares its layers (i.e., hard parameter sharing) to learn a generic representation which is then used as input of independent task-specific decoder heads. 

Newly, a single segmentation network \cite{boutillon2022generalizable} proposed to manage multiple inputs/outputs along with intra-domain shift by introducing contrastive regularization at shared representations levels. In the meantime, semi-overcomplete shape priors \cite{sadikine2022semi}, encoded by a deep semi-overcomplete convolutional auto-encoder, were incorporated into a vessel segmentation pipeline to efficiently manage shape heterogeneity. In addition, a hierarchical progressive multi-scale network was developed in \cite{hao2022hpm} to exploit different receptive field sizes, combined with a deep supervision mechanism to accelerate network convergence. Despite these advances, multi-scale geometry, as an intrinsic characteristic of vascular trees, has been weakly investigated in the literature. This motivated us to design a new approach providing a strong focus on the different scales arising from vascular networks.

In this work, our contributions are two-fold. First, we propose a multi-scale clustering methodology allowing liver vasculature decomposition relying on inherent statistics and estimated branch radius of the vascular tree. Second, a newly designed end-to-end deep multi-task segmentation framework (Fig.\ref{fig:fig1}) with multi-scale contrastive learning \cite{oord2018representation, chen2020simple} is developed with the ability to discriminate intra-scale compactness and inter-scale separability at different scale levels. The effectiveness of our vascular segmentation pipeline based on scale-specific auxiliary multi-task contrastive learning is illustrated for liver vessel extraction in the 3D-IRCADb \cite{soler20103d} dataset.

\section{Methods}
\label{sec:methods}

Deep learning-based liver vessel segmentation from medical images is addressed in a multi-task fashion, relying on scale-specific auxiliary sub-tasks. Let us denote $\mathcal{D}=\{(\pmb{x}_{i},\{\pmb{y}_{i}^{s}\}_s)\}$ as the training dataset where $s\in \{0,..., S\}$ indexes task referring to different scales, $\pmb{x}_i$ a greyscale volume and $\{\pmb{y}^{s}_i\}_{s=0}^{S}$ the multiple sets of ground truth vasculature segmentation masks, with voxels coordinates $\nu$. In our case, we define $s$ as a specific vascular scale that belongs to the binary mask $\pmb{y}= \cup_{s=1}^{S}\pmb{y}^{s}$ where $\pmb{y}^{0}$ reflects the initial scale (i.e., $\pmb{y}=\pmb{y}^0$). In this setting, supervised multi-task segmentation with a single input, multiple outputs and $S$ learning tasks $\{\mathcal{T}_{s}\}$ consists of approximating a mapping function $\xi:\pmb{x}\rightarrow \xi(\pmb{x})=\{\hat{\pmb{y}}^{s}\}_{s=0}^{S}$ from $\mathcal{D}$, through a deep multi-task convolutional network.

\subsection{Multi-scale vessel clustering}
\label{ssec:clustering}

Defined as a set of $n_b$ branches $\{B_j\}_{j=1}^{n_b}$ in the binary segmentation mask $\pmb{y}$, vessels are assumed to be tubular, which insinuates that each branch $B_j \subset \pmb{y}$ is regular and has a constant radius $r_j$. This assumption is denoted as hypothesis $\pmb{H}$. To estimate all the radii $\{r_j\}_{j=1}^{n_b}$ at the scale of the entire vascular network, both edge surface (noted as $\Omega$) detection and thinning \cite{lee1994building} algorithms were sequentially applied (Fig.\ref{fig:fig2}\textit{a}). The latter extracts a voxel level skeleton $\Gamma \subset \pmb{y}$, with a set of voxels coordinates $\{\gamma_k\}^{q}_{k=1} \subset \nu$. Based on this, a local radius $\hat{\sigma}_{k}$ estimated in a given orthogonal cross-section of $\pmb{y}$ with centroid $\gamma_k$ is derived from the nearest $m$-tuplet voxel neighbors $\{\omega^{p} \hspace{0.1cm}| \hspace{0.1cm}\omega^{p-1} < \omega^{p}\}_{p=1}^{m}$ with $\omega^{p} = \arg \min_{\omega \in \Omega} \|\gamma_{k}-\omega\| \in \Omega$ the distance to the closest vessel surface voxel, as follows:

\vspace{-0.25cm}
\begin{equation}
    \hat{\sigma}_{k}=\max_{p \leq m}\,\|\gamma_{k}-\omega^p\|
\label{eq:eq1}
\end{equation} 
\vspace{-0.25cm}

To estimate the radii $\hat{r}_j$ from the set of local radii $\{\hat{\sigma}_{k}\}^{q}_{k=1}$, we cluster in an unsupervised manner the skeleton branches into $n_b$ classes (i.e., each class in the skeleton $\Gamma$ is assigned as a branch of index $j$). The clusterized skeleton is noted as $\Gamma^{\prime}$, beneath the assumption $\pmb{H}$:

\vspace{-0.25cm}
\begin{equation}
    \hat{r}_{j}=\median_{k\leq q}\{\hat{\sigma}_{k} \hspace{0.1cm} | \hspace{0.1cm} \gamma_k \in B_j\} 
\label{eq:eq2}
\end{equation}
\vspace{-0.25cm}

\input{fig2.tex}

To reconstruct the labeled branches $\{B_j\}_{j=1}^{n_b}$, we rely on the Euclidean Distance Transform (EDT), defined for a given branch as in Eq.\ref{eq:eq3}. Thus, we look for the voxel $\nu$ closest to $\gamma^{\prime j}_k$ (i.e., the voxel of the centerline labeled by the index of the branch $j$) with $\nu$ not being part of the centerline. This means finding the voxels orthogonal to $\gamma^{\prime j}_k$ and labeling them as part of the branch $B_j$.

\vspace{-0.25cm}
\begin{equation}
    B_j(\nu)=j \, \hspace{0.2cm}\textrm{s.t.}\, \min_{\nu \notin \Gamma^{\prime}}\,\|\gamma^{\prime j}_k-\nu\| \,\hspace{0.2cm} \forall k \,\hspace{0.1cm} \textrm{and} \,\hspace{0.1cm} \pmb{y}(\nu)=1
\label{eq:eq3}
\end{equation}
\vspace{-0.25cm}

As a final step, we define $S-1$ thresholds based on the intrinsic statistics of each volume $\pmb{y}_i$. This choice is related to the intra-volume variability in order to generate $S$ ground truth masks $\{\pmb{y}_{i}^{s}\}$, each corresponding to a given scale $s$. In our study, we fixed $S=3$, which is equivalent to decomposing the vascular tree into large, medium, and small blood vessels (Fig.\ref{fig:fig2}\textit{b}). In this context, the first and the second threshold were estimated from $\{\hat{r}_{j}\}_{j=1}^{n_b}$ and the estimators were defined as the first quartile $Q_1$ and the third quartile $Q_3$ (Fig.\ref{fig:fig4}).

\input{fig3}

\subsection{Multi-Task Learning (MTL)}
\label{ssec:multi}

To approximate the mapping function $\xi$, a multi-task convolutional network with single input and multiple outputs is employed (Fig.\ref{fig:fig1}). It comprises a contracting path encoder $E=E^{2^lf_0} \circ \cdots \circ E^{f_0}$ dealing with a succession of convolutions, Batch Normalization (BN), non-linearity, and max-pooling, which projects $\pmb{x}$ to a latent representation denoted as $\pmb{z}$ (Fig.\ref{fig:fig3}), where $l$ and $f_0$ respectively designates the number of hidden layers and the initial number of features maps. The developed architecture also includes multiple expansive pathway decoders $D_s=D^{f_0}_s \circ \cdots \circ D^{2^{(l-1)}f_0}_s $ composed of a cascade of up-sampling, long symmetrical skip-connections, convolutions, BN, and non-linearity, except $D^{f_0}_s$ which comprises an additional convolution with $1\times1\times1$ kernel as well as an activation function (Fig.\ref{fig:fig3}). Each decoder projects back the shared representation $\pmb{z}$ to achieve a task-specific segmentation map $\hat{\pmb{y}}^{s}$. This conducted us to define in a general way the network (denoted $\xi$ as the mapping function it approximates) with numerous tasks $\{\mathcal{T}_{s}\}$, following:

\begin{equation}
    \xi(\pmb{x})=\{D_s(\pmb{z})=\hat{\pmb{y}}^{s}\hspace{0.1cm}|\hspace{0.1cm} \pmb{z}=E(\pmb{x})\}_{s=0}^{S}
\label{eq:eq4}
\end{equation}
\vspace{-0.25cm}

\noindent where $D_0(\pmb{z})=D_{mt}(\pmb{z})$ and $\{D_s(\pmb{z})\}_{s=1}^{S}$ refer respectively to the main ($mt$ stands for main task) and the auxiliary tasks decoders. Furthermore, the model parameters are estimated by learning both main and auxiliary tasks jointly through the optimization of the following loss function:

\vspace{-0.25cm}
\begin{equation}
    \mathcal{L}=\mathcal{L}_{mt}(\pmb{y},\hat{\pmb{y}})+\sum_{s=1}^{S}\lambda_s\mathcal{L}_{s}(\pmb{y}^{s},\hat{\pmb{y}}^{s})
\label{eq:eq5}
\end{equation}
\vspace{-0.25cm}

\noindent where $\mathcal{L}_{mt}$ and $\mathcal{L}_{s}$ combine Dice and weighted cross-entropy. Hyper-parameters $\lambda_s$ balance the contributions of each scale-specific auxiliary task.

\vspace{-0.25cm}

\subsection{Contrastive multi-task multi-scale learning}
\label{ssec:contrastive}

All tasks shared the same representation $\pmb{z}$, which may bring confusion to the encoder $E$. To handle this, in favor of the main task, constraining $E$ to manage inter- and intra-auxiliary task relationships is needed. Therefore, we propose to add an additional loss to Eq.\ref{eq:eq5}, allowing to benefit from contrastive learning \cite{chen2020simple}. Defined as a hardness-aware loss function, the contrastive loss encourages features representation $\pmb{z}^{s}_i$, belonging to an anchor set $\mathcal{A}$ from the same scale to be aligned (i.e., closeness) and separate features from different scales apart in order to ensure uniformity (i.e., uniform distribution in a hypersphere). Since it is difficult to make the distinction between the latent representation of each auxiliary task at $\pmb{z}$, we define the compact representation derived from the first layer of the decoder $D_s$ for each task $\mathcal{T}_{s}$, following:

\begin{equation}
    \pmb{z}_{i}^{s}=\proj_{s}(D^{2^{(l-1)}f_0}_s)
\label{eq:eq6}
\end{equation}
\vspace{-0.25cm}

\noindent where $\proj_{s}$ applies a sequence of $1\times1\times1$ convolutions, BN and non-linearity (Fig.\ref{fig:fig3}) to map the dense feature $D^{2^{(l-1)}f_0}_s$ to a lower dimensional space. The representations $\pmb{z}^{s}_i$ from the same scale (resp. different scale) belong to the set $\mathcal{P}_i$ (resp. $\mathcal{N}_i$). From this, the contrastive loss can be written as:

\begin{equation}
    \mathcal{L}_{c}=\frac{1}{|\mathcal{A}|} \sum_{i \in \mathcal{A}} \frac{1}{|\mathcal{P}_i|} \sum_{j \in \mathcal{P}_i} l_{i,j}
\label{eq:eq7}
\end{equation}
\vspace{-0.25cm}

\noindent where:
\vspace{-0.25cm}
\begin{equation}
    l_{i,j}=- \log \frac{\exp ({\pmb{z}_i^s}^\top\pmb{z}_j^s / \tau)}{\exp ({\pmb{z}_i^s}^\top\pmb{z}_j^s  / \tau)+\sum_{k\in\mathcal{N}_i}\exp ({\pmb{z}_i^s}^\top\pmb{z}_k^s / \tau)}
\label{eq:eq8}
\end{equation}

\noindent $\pmb{z}^s$ is a $l_2$-normalized vector and $\tau > 0$ a scalar temperature hyper-parameter. Finally, this leads us to propose Eq.\ref{eq:eq9} as global loss function:

\vspace{-0.25cm}
\begin{equation}
    \mathcal{L}_{total} = \mathcal{L}_{mt}+\sum_{s=1}^{S}\lambda_s\mathcal{L}_{s} + \lambda_{c}\mathcal{L}_{c}
\label{eq:eq9}
\end{equation}
\vspace{-0.25cm}

\noindent where $\lambda_{c}$ depicts a hyper-parameter that regulates the strength of contrastive learning.

\input{tab.tex}
\section{Experiments}
\label{sec:experiment}

\subsection{Imaging datasets}

We assess the proposed approach on the publicly-available abdominal 3D-IRCADb \cite{soler20103d} dataset containing CT scans with ground truth masks from 20 patients (10 women, 10 men) with liver tumours in $75\%$ of cases. After extracting a liver bounding box in each CT scan, we resized all volumes to $256\times256\times128$ voxels with isotropic size. Then, we performed the clustering of ground truth masks described in Sect.\ref{ssec:clustering} with $p=8$ to generate $\{\pmb{y}_{i}^{s}\}$.

\input{fig4}

\subsection{Implementation details}
For all experiments, the number of layers $l$, the number of features maps $f_0$, the learning rate, batch size, and the number of epochs were respectively set to $5$, $8$, $3\times10^{-4}$, $2$, and $2000$. After hyper-parameters mining using Optuna \cite{optuna_2019}, optimal $\lambda_1$, $\lambda_2$, $\lambda_3$, $\lambda_c$ and $\tau$ (Eq.\ref{eq:eq9}) were empirically set to $0.78$, $0.48$, $0.54$, $0.53$ and $0.94$. Random data augmentation was involved on the fly during training: rotation, translation, flipping, and gamma correction. 5-fold cross-validation was followed, and segmentation networks were implemented with PyTorch. Seeds were fixed for weight initialization, data augmentation, and data shuffling to ensure reproducibility.

\subsection{Evaluation of predicted segmentation}

To evaluate the performance of our model against existing approaches, we compared ground truth $\mathit{GT}$ and predicted $\mathit{P}$ masks through the following metrics: Dice coefficient \texttt{DSC} ($\frac{2|\mathit{GT}\cap \mathit{P}|}{|\mathit{GT}|+|\mathit{P}|}$), Jaccard coefficient \texttt{Jacc} ($\frac{|\mathit{GT}\cap\mathit{P}|}{|\mathit{GT}\cup\mathit{P}|}$) as well as \texttt{clDSC} coefficient \cite{shit2021cldice} for connectivity assessment. The Hausdorff distance $\textstyle{\mathit{\HD(\mathit{GT}, \mathit{P})}=\max(\mathit{h(\mathit{GT}, \mathit{P})}, \mathit{h(\mathit{P}, \mathit{GT})})}$ was also used with $\textstyle{\mathit{h(\mathit{A}, \mathit{B})}= \max_{\mathit{a} \in \mathit{A}}\min_{\mathit{b} \in \mathit{B}}\,\|\mathit{a}-\mathit{b}\|}$.

\section{Results and discussion}

Regarding multi-scale vessel clustering, the letter-value plot in Fig.\ref{fig:fig4} reveals a substantial heterogeneity in intra-volume $\pmb{y}_i$ multi-scale geometry (i.e., shape, topology) within the 3D-IRCADb dataset. Moreover, this finding explains the intrinsic statistics choice per volume highlighted in Sect.\ref{ssec:clustering}. This concretely demonstrates the difficulties that a deep network can face in such a challenging liver vessel segmentation task.

Concerning the segmentation, the baseline 3D ResUNet models (i.e., $S=0$ in Eq.\ref{eq:eq4}) in binary and multi-class settings were compared with the proposed method, consisting of multi-task learning with contrastive multi-scale auxiliary sub-tasks. The quantitative results from Tab.\ref{tab:qresults} demonstrate, under the assumption of reaching the optimal loss function hyper-parameters, that our approach outperforms the baseline in terms of \texttt{DSC}, \texttt{Jacc} and \texttt{clDSC} with a gain of $0.79\%$, $0.85\%$, and $1,29\%$ respectively, except for \texttt{HD} where the binary segmentation baseline overwhelms our method. On the other hand, the model performance drops out in multi-class setting, which can be explained by the inter-scale confusion faced by the model in its hidden layers. Moreover, the ablation study demonstrates the effectiveness of integrating multi-scale contrastive learning into Eq.\ref{eq:eq5}, especially in \texttt{clDSC}. Further, Fig.\ref{fig:fig5} illustrates the connectivity improvement reached by our method. Ultimately, it is worth mentioning that during the inference stage, only the main task decoder $D_{mt}$ (Fig.\ref {fig:fig1}) was employed, which is desirable towards the deployment of our pipeline into clinical routine.

\input{fig5}

\section{Conclusion}
\label{sec:conclusion}

We have unveiled a novel approach for liver vessel segmentation based on scale-specific auxiliary multi-task contrastive learning. In future works, it would be valuable to evaluate the proposed method with a various number of vascular tree scales to more deeply assess its robustness. In addition, incorporating shape and topological priors could allow the model to avoid prediction confusion and increase its generalization ability. Furthermore, a memory bank storing multi-scale latent representations could be a way to further benefit from contrastive learning while alleviating batch size constraints. More globally, our contributions could be easily extended to other types of vasculature including brain and pulmonary vessels from diverse imaging modalities.

\section{Compliance with ethical standards}
\label{sec:ethics}

This research study was conducted retrospectively using human subject data made available in open access \cite{soler20103d}.

\bibliographystyle{IEEEbib}
\bibliography{References}

\end{document}

%% file: fig1.tex
\begin{figure}[!t]
    \centering
    \includegraphics[width=\linewidth]{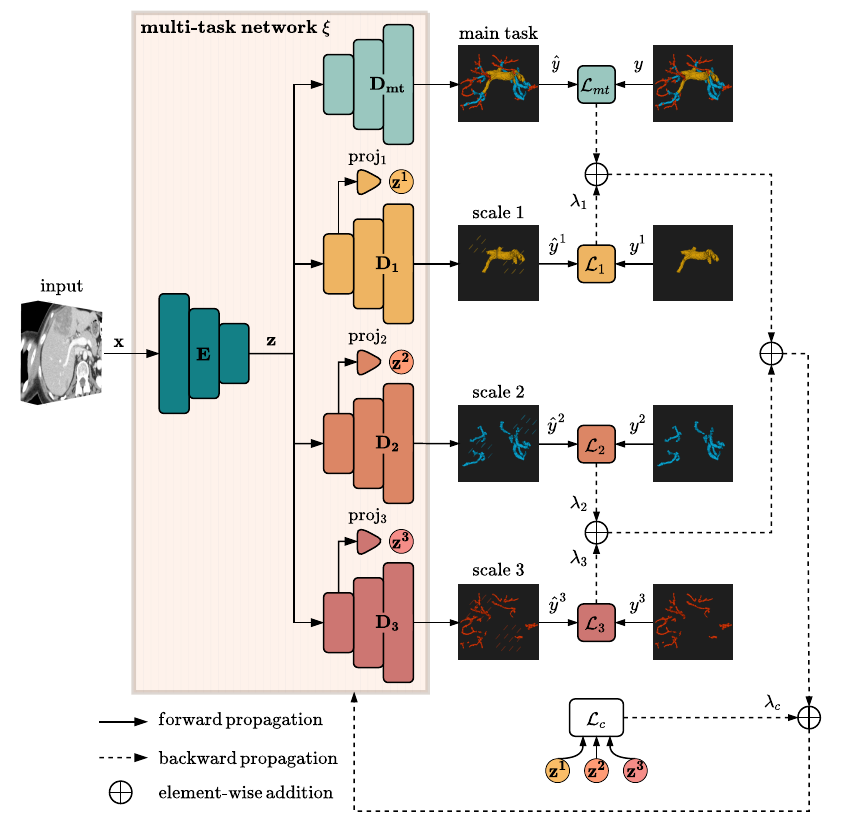} \vspace{-0.5cm} \\
    \caption{Proposed pipeline for deep liver vessel segmentation using scale-specific auxiliary multi-task contrastive learning.} \vspace{-0.2cm}
    \label{fig:fig1}
\end{figure}

%% file: fig2.tex
\begin{figure}[t!]
\centering
\begin{minipage}[b]{0.6\columnwidth}
  \centering
  \centerline{\includegraphics[width=\linewidth]{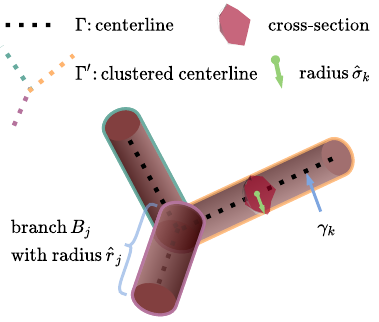}}
  \centerline{(a)}\medskip
\end{minipage}
\begin{minipage}[b]{0.35\columnwidth}
  \centering
  \centerline{\includegraphics[width=\linewidth]{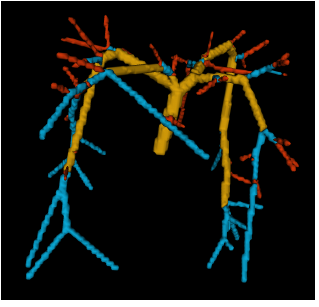}}
    \vspace{.3cm}
  \centerline{(b)}\medskip
\end{minipage} \vspace{-0.2cm}
\caption{(\textit{a}) Visual aid to interpret notations related to multi-scale vessel clustering (Sect.\ref{ssec:clustering}). (\textit{b}) Example of 3-scale vasculature clustering applied on a synthetic vasculature \cite{jassi2011vascusynth}.} \vspace{-0.05cm}
\label{fig:fig2}
\end{figure}

%% file: fig3.tex
\begin{figure}[!t]
    \centering
    \includegraphics[width=\linewidth]{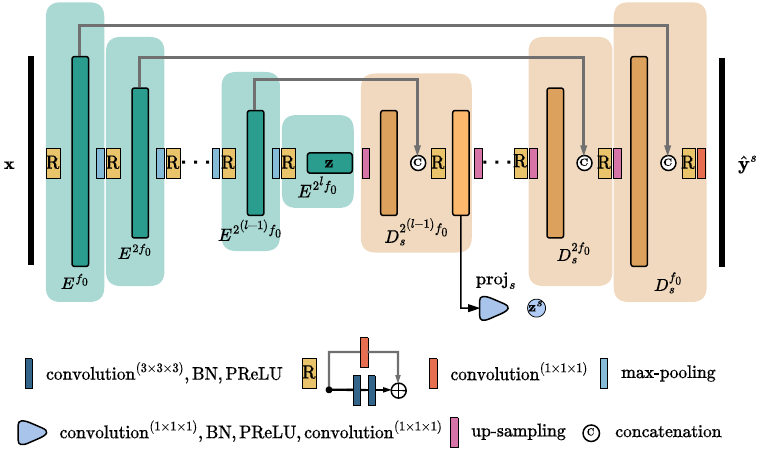}
    \caption{Encoder-decoder architecture involved in our multi-task contrastive pipeline. PReLu stands for Parametric ReLU.} \vspace{-0.2cm}
    \label{fig:fig3}
\end{figure}

%% file: tab.tex
\begin{table*}[!t]
\centering
\vspace{-0.25cm}
\caption{Liver vessel segmentation on CT scans from the 3D-IRCADb \cite{soler20103d} dataset using 3D ResUNet in binary and multi-class settings as well as the proposed approach without and with contrastive learning. Best results are displayed in bold.}
\resizebox{1.3\columnwidth}{!}{%
\begin{tabular}{@{}llcccc@{}}\cmidrule(l){3-6} 
\multicolumn{2}{c}{Models} &
  \shortstack{\texttt{DSC}$\uparrow$ \\ score ($\%$)} &
  \shortstack{\texttt{Jacc}$\uparrow$ \\ score ($\%$)} &
  \shortstack{\texttt{clDSC}$\uparrow$ \\ score ($\%$)} &
  \shortstack{\texttt{HD}$\downarrow$\\ dist. ($\mathrm{mm}$)} \\ \midrule
\multirow{2}{*}{3D ResUNet}   & binary          & $54.25\pm3.65$       & $37.65\pm3.46$       & $46.84\pm3.63$       & $\pmb{80.09}\pm4.83$ \\
                               & multi-class     & $50.31\pm4.43$       & $34.13\pm3.98$       & $44.34\pm2.89$       & $114.3\pm10.9$      \\ \midrule
\multirow{2}{*}{\textbf{Ours}} & $\lambda_c=0$   & $54.27\pm4.16$       & $37.73\pm3.87$       & $46.90\pm2.44$       & $86.07\pm8.07$       \\
                               & $\lambda_c\ne0$ & $\pmb{55.04}\pm2.79$ & $\pmb{38.50}\pm2.71$ & $\pmb{48.13}\pm3.52$ & $82.36\pm12.3$      \\ \midrule
                               &                 & \multicolumn{1}{l}{} & \multicolumn{1}{l}{} & \multicolumn{1}{l}{} & \multicolumn{1}{l}{}
\end{tabular}%
}
\vspace{-0.6cm}
\label{tab:qresults}
\end{table*}

%% file: fig4.tex
\begin{figure}[!ht]
    \centering
    \includegraphics[width=1\columnwidth]{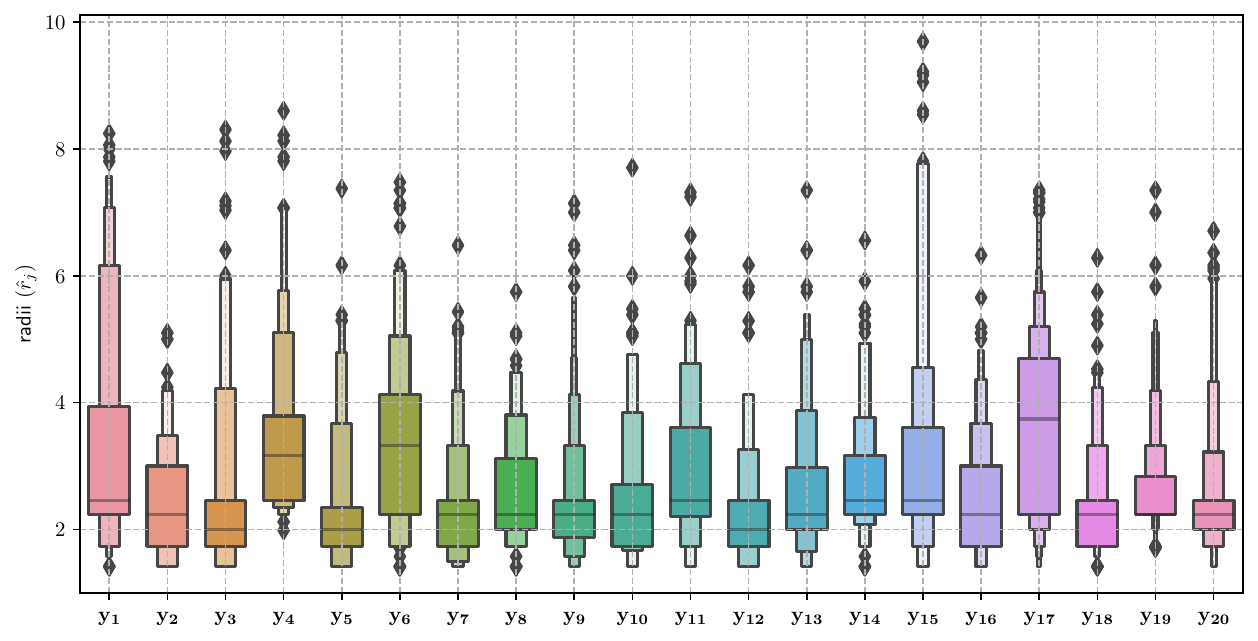}
    \caption{Letter-value plot of branch radii $\hat{r}_{j}$ $(\mathrm{mm})$ per ground truth volume $\pmb{y}_i$ from the 3D-IRCADb \cite{soler20103d} dataset.} \label{fig:fig4} \vspace{-0.2cm}
\end{figure}

%% file: fig5.tex
\begin{figure}[!t]
    \centering
    \includegraphics[width=1\columnwidth]{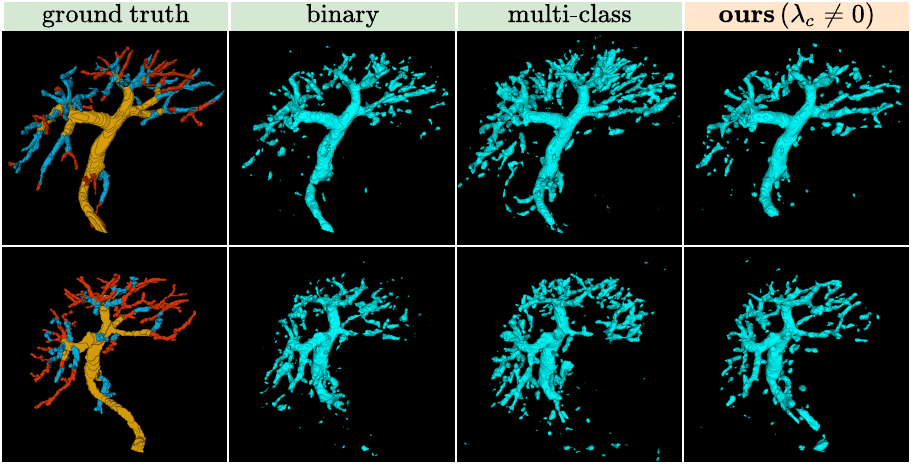}
    \caption{Qualitative liver vessel segmentation results on CT scans from the 3D-IRCADb \cite{soler20103d} dataset.}
    \label{fig:fig5}
    \vspace{-0.2cm}
\end{figure}